\def\inArXiv{defined}
\def\inDCC{defined}
  \ifx\inIEEE\undefined
    \ifx\inDCC\undefined
      \documentclass[10pt,a4paper]{article} 
      \usepackage[authoryear]{natbib}
      \bibpunct{(}{)}{;}{a}{}{,}
    \else
      \documentclass[smallcaptions,smallabstract]{dccpaper} 
      \usepackage[numbers]{natbib}
      \ifx\inDCCtweaked\undefined
      \else
        \usepackage[right=1in,left=1in,top=1in,bottom=1in]{geometry}
      \fi
      \makeatletter
      \renewcommand{\@biblabel}[1]{[#1]}
      \makeatother
      \makeatletter
      \g@addto@macro\normalsize{%
        \setlength\abovedisplayskip{0.5em}
        \setlength\belowdisplayskip{0.5em}
        \setlength\abovedisplayshortskip{-0.1em}
        \setlength\belowdisplayshortskip{0.5em}
      }
      \g@addto@macro\small{%
        \setlength\abovedisplayskip{0.4em}
        \setlength\belowdisplayskip{0.4em}
        \setlength\abovedisplayshortskip{-0.1em}
        \setlength\belowdisplayshortskip{0.4em}
      }
      \makeatother
    \fi
  \else
    \documentclass[10pt,journal]{../style/IEEEtran}  

    \ifx\useIEEEbib\undefined
      \usepackage[authoryear]{natbib}
    \else
      \usepackage[numbers]{natbib}
    \fi
  \fi
\usepackage[utf8]{inputenc}
\usepackage[russian,english]{babel}
\ifx\inDCC\undefined
  \usepackage{fancyhdr}
\fi
\def\papertitle{Compressing combinatorial objects}
\def\paperauthors{Christian Steinruecken}

\DeclareFontShape{OT1}{cmtt}{bx}{n}{<5><6><7><8><9><10><10.95><12><14.4><17.28><20.74><24.88>cmttb10}{}
\usepackage[pdfnewwindow=true,breaklinks,bookmarks=false]{hyperref}
\usepackage{breakurl}
\usepackage{etoolbox}
\usepackage{tikz}
\usetikzlibrary{shapes,shadows,arrows,calc}
\usepackage{theorem}
\usepackage{graphicx}
\usepackage{subfig}
\usepackage{placeins}
\usepackage{afterpage}
\usepackage{makeidx}
\usepackage{booktabs}
\usepackage{dashrule}
\usepackage{listings}
\usepackage{msyntax}
\usepackage{mthesis}

\definecolor{darkpurple}{cmyk}{0.0,1.0,0.25,0.25}%
\hypersetup{
    pdftitle={\papertitle},    
    pdfauthor={\paperauthors},     
    pdfcreator={\LaTeX, dvips, ps2pdf},   
    pdfkeywords={Data compression, arithmetic coding, permutations, combinations, multisets, combinatorics},
    pdfnewwindow=true,      
    bookmarks=false,
    colorlinks=true,
    pdfborder={0 0 0},
    citecolor=,
    linkcolor=,
    urlcolor=blue,
    filecolor=magenta,
}
\begin{document}
\def\pieceofwriting{article}
\def\segment{article}

\def\Apndxname{Appendix}%
\def\apndxname{appendix}%
\def\Apndcsname{Appendices}%
\def\apndcsname{appendices}%


\newif\ifIEEE
\ifx\IEEEkeywords\undefined
  \IEEEfalse
\else
  \IEEEtrue
\fi

\ifx\inDCC\undefined
  \def\thetitle{\papertitle}
\else
  \def\thetitle{\textbf{\papertitle}}
\fi

\def\algname{Algorithm}
\def\Algname{Algorithm}

\newif\iftwocolumn
\makeatletter
\if@twocolumn
  \twocolumntrue
\else
  \twocolumnfalse
\fi
\makeatother

\ifx\inDCC\undefined
  \def\maybesmall{}
  \def\maybedense{}%
  \def\DCCvspace#1{}
\else
  \def\maybesmall{\renewcommand{\baselinestretch}{0.925}%
                  \small}
  \def\maybedense{\setlength{\itemsep}{-0.125cm}}%
  \def\DCCvspace#1{\vspace*{#1}}
  %
  \makeatletter
  \if@smallcaptions
    \newsavebox{\dcctempbox}
    \renewcommand{\@makecaption}[2]
    {\vspace{10pt}\renewcommand{\baselinestretch}{\smallstretch}
     \small\sbox{\dcctempbox}{#1: #2}
     \ifthenelse{\lengthtest{\wd\dcctempbox > \linewidth}}
     { #1: #2\par}{\begin{center}#1: #2\end{center}}}
  \fi
  \makeatother
\fi

\title{\large \thetitle}
\ifx\inIEEE\undefined 
  \author{{Christian Steinruecken}\\ {\small\hspace*{-1.7em} University of Cambridge}}
\else 
  \author{Christian Steinruecken
  }
\fi
\maketitle

\vspace*{2\bigskipamount}

%
\def\ctx#1{\seqvar{#1}}%
\def\suf#1{\text{suf}(#1)} 
\def\pars{hyper-parameters}%
\def\depfan{\fan\dep}%
\def\DEPFAN{\FAN\times\DEP}%

\def\n[#1]{\ensuremath{n_{\sym{#1}}}}%
\def\TT{\mbox{\tiny\textsf{T}}}%
\def\nt{\ensuremath{n_{\TT}}}%

\makedistvar{DD}{D}%
\makedistvar{PP}{P}%
\def\txtlogJJ{\ensuremath{\log_2 \txtJJ}}%
\makemultiset{M}{m}{M} 
\def\mk{\mm{x}}
\def\xk{x}%
\def\Mbot{x \in \setX}%
\def\Mtop{}%
\makemultiset{W}{w}{W} 
\makemultiset{R}{r}{R} 
\def\LLomit#1#2{\LL[\setminus#1]{#2}}%
\def\esc{\texttt{ESC}}%
\def\Ksize{$K$\kern-.12em-size}
\def\fcnt#1#2#3{\sum_{#2}^{#3} \one{#1}}%
\def\thesisChapter#1{chapter~#1 of \citep{steinruecken2014b}}%
\def\ThesisChapter#1{Chapter~#1 of \citep{steinruecken2014b}}%
\def\thesisChapter#1{\citep{steinruecken2014b5}}%
\makeatletter
\def\@ctxnum[#1]#2{\ensuremath{n^{#1}_{#2}}}
\def\@ctxNUM[#1]{\ensuremath{N_{#1}}}
\def\@ctxUNUM[#1]{\ensuremath{U_{#1}}}
\def\@ctxTNUM[#1]{\ensuremath{T_{#1}}}
\def\num{\@testopt{\csname @ctxnum\endcsname}{\kern.2ex}}
\def\NUM{\@testopt{\csname @ctxNUM\endcsname}{\kern.2ex}}
\def\UNUM{\@testopt{\csname @ctxUNUM\endcsname}{\kern.2ex}}
\def\TNUM{\@testopt{\csname @ctxTNUM\endcsname}{\kern.2ex}}
\makeatother

%
%
%
%

\makeatletter%
\def\declareHumNN#1#2#3{%
  \expandafter\def\csname #1\endcsname{\@humbleformat{#2}#3}%
  \expandafter\def\csname s#1\endcsname{{#2}#3}%
  \expandafter\def\csname idx#1\endcsname{{\@idxhumbleformat{#2}#3}}%
}%
\makeatother%
\def\vp{$^+$}%
\declareHumNN{yaN}{YA}{9}%
\declareHumNN{yaNP}{YA}{9\vp}%
\declareHumNN{ycS}{YC}{16}%
\declareHumNN{ycSP}{YC}{16\vp}%
\declareHumNN{naEM}{N}{8}%
\declareHumNN{naEP}{N}{8\vp}%
\declareHumNN{naSM}{N}{16}%
\declareHumNN{naSP}{N}{16\vp}%
%
\declareHumble{PPMDP}{PPM\raise0.24ex\hbox{-}DP}

%
\def\componentwise{com\-po\-nent-wise}%
\def\nobreakhyphen{\raise0.1ex\hbox{-}}%
\def\SHA#1{SHA\nobreakhyphen{}\kern-0.1ex#1}%

%
%
%

\ifx\inthesis\undefined
\fi

\index{multisets|(}%

%


\begin{abstract}
%
%
Most of the world's digital data is currently encoded
in a sequential form,
and compression methods for sequences have been studied
extensively.
%
However, there are many types of non-sequential data
for which good compression techniques are still largely unexplored.
This paper contributes insights and concrete techniques for compressing
various kinds of non-sequential data via arithmetic coding,
and derives re-usable probabilistic data models from fairly generic
structural assumptions.
Near-optimal compression methods are described for
certain types of permutations, combinations and multisets;
and the conditions for optimality are made explicit for each method.
%

%
%
%
%
%
\end{abstract}

\ifx\IEEEkeywords\undefined
\else
\fi

\section{Introduction}


Much of data compression is concerned with the task of
compressing files that commonly occur in every day usage.
Improvements in compression effectiveness inevitably stem from
improvements in data modelling, and the strength of any given
model clearly depends on the type of data that is being
compressed.
%
The need for good compression thus motivates developing models that are
good at predicting the data of interest.
The development of appropriate data models can be difficult,
especially when computational constraints are taken into account.
For example, models that are good at compressing \DNA\ sequences
are different from models that are good at compressing human text.
Modern file compressors often include many different data models
(often for specific types of data) whose predictions are combined
\citep{mahoney2000a,mahoney2002a,mahoney2005a}.
%

%
%

Sequence compression algorithms have been studied extensively,
not least because sequences are the most common form in which
data is represented in a computer.
%
%
%
%
%
Of course, a~serial
representation is not always a natural description of data
(e.g.~2-dimensional pictures, elements of a dictionary, the
hierarchical structure of a file system),
and is often chosen somewhat arbitrarily.

This paper explores data models (and compression algorithms) for
some basic types of data that are non-sequential, or have
particular symmetries imposed on them.  In particular,
various models
are proposed for compressing different kinds of permutations,
combinations, and multisets.


The methods for compressing data with these structures
can be used as components in larger models and compression algorithms;
we also hope that the insights presented for the objects here
may be useful for finding algorithms for other kinds of
combinatorial objects.
%
%

\bigskip

The rest of this paper is structured as follows.
\def\bb#1{\textit{#1}}%
\bb{Section~\ref{sec:iid-seq}} reviews the compression
of sequences whose elements are independent and identically
distributed from a known distribution.
\bb{Section~\ref{sec:iid-ms}} introduces multisets and describes
an optimal compression scheme for multisets whose
elements are distributed the same way as in section~\ref{sec:iid-seq}.
\bb{Section~\ref{sec:perm}} introduces permutations and
describes an optimal compression scheme for
permutations that are uniformly distributed.
\bb{Section~\ref{sec:equality}} shows that a sequence can
be split into a multiset and a permutation, and that
the information content of these two components
sum to the same value as the information content
of the original sequence.
\bb{Section~\ref{sec:truncperm}} extends the coding
method for permutations to truncated permutations.
\bb{Section~\ref{sec:comb}} introduces combinations
and a compression scheme for them.
\bb{Section~\ref{sec:uniform-ms}} discusses
uniformly distributed multisets and describes
a compression scheme for them.
%
%
Finally, \bb{Section~\ref{sec:learn}} extends the results of
previous sections
to the case that the symbol distribution is
unknown and needs to be learned adaptively.


\section{Sequences}\label{sec:iid-seq}

We'll start by reviewing a task that is well known:
the compression of a fixed-length sequence
$\vc{x}{1}{N}$ of elements from a finite alphabet $\setX$.
Rather than discussing a specialised model that is good at compressing
text or other real-world sequences,
we'll look at an oversimplified sequence model
and then show equivalents for non-sequential forms of data.
%
So let's assume for now that each element $x_n$ in the sequence is
independently and identically distributed (\iid)
according to some distribution $\txtDD$.
%
The probability of such a sequence is:
\begin{equation}\label{eq:iidseq}
\p{\vc{x}{1}{N} \| \txtDD, N}
  =  \prod_{n=1}^N \DD{x_n}     \maybedot
\end{equation}
%
Sequences of this kind can be compressed with an arithmetic
coding scheme \citep{witten1987a,moffat1998a},
where the symbols $x_n$ are encoded sequentially
using $\txtDD$ at each coding step.%
\footnote{%
  There are some
  fairly minor constraints, e.g.~that it must be possible
  to discretise~$\txtDD$ to the
  resolution of the arithmetic coder.
}
This kind of algorithm produces compressed output
whose total length is guaranteed to be within 2~bits
of the theoretic optimum --
the \emph{Shannon information content} 
of the sequence
under the above assumptions:
%
\begin{equation}\label{eq:ic-iid-sequence}
        \log_2 \frac{1}{\prod_{n=1}^N \DD{x_n}}
  \ =\  \sum_{n=1}^N \log_2 \frac{1}{\DD{x_n}}
\maybedot
\end{equation}%
This guarantee is strictly stronger than
proving
the \emph{expected} output length to be
within 2~bits of the Shannon entropy of $\txtDD$,
because the guarantee holds
for every input sequence, not just on average.
Algorithms of the above kind are well known,
and form the basis of many sequence compression algorithms.%
\footnote{%
  Of course, there are many ways of modelling sequences,
  and the above simplistic model, while fundamental, is hardly
  suitable for compressing real-life files.
  In fact, if $\txtDD$ were chosen to be a uniform distribution,
  the above method achieves no compression at all, as
  all sequences of length $N$ are assumed to have equal probability.
}
%

\medskip

%
%

This paper now explores similarly fundamental compression
models for input objects that \emph{aren't} sequences.
Compression of non-sequential data was considered by
\citet{varshney2006b,varshney2006a,varshney2007a},
and this paper follows up on their work.
The models in this paper have been chosen primarily
for simplicity and clarity of presentation;
more elaborate models for combinatorial objects
can be found e.g.~in~\thesisChapter{5}.


%



\section{Multisets}\label{sec:iid-ms}

A~\emph{multiset} $\setM$ is an unordered collection of elements,
where elements may occur multiple times.
Let $\mm{x}$ denote how often element $x$ occurs in $\setM$,
and $\MM$ the total number of elements in $\setM$ (including
repetitions).
For example, $\setK = \bset{\sym{A},\sym{B},\sym{B}}$
is a multiset with
$\KK=3$, $\kk{\sym{A}}=1$, and $\kk{\sym{B}}=2$.
%

One can think of a multiset as the \emph{histogram} 
that captures the symbol occurrence counts in a sequence $\vc{x}{1}{N}$,
such that:
\begin{equation}
\MM \ =\  \sum_{\Mbot}^{\Mtop} \mk
    \ =\  N
    \maybedot
\end{equation}
%
A multiset represents the information that remains when
the order information of a sequence is lost.

%

Analogously to how we considered a sequence of $N$ draws in the
previous section,
consider now
a multiset $\setM$ that was created by making $N$ independent
draws from a known distribution $\txtDD$
over the same finite alphabet $\setX$.
%
The probability of such a multiset~$\setM$ is:
\begin{equation}\label{eq:iid-ms}
\p{\setM \| N, \txtDD} = N! \mul \prod_{\Mbot}^{\Mtop}
      \frac{{\DD{\xk}}^{\mk}}{\mk!}
    \maybedot
\end{equation}
%
%
As $\setM$ is a multivariate object, the above distribution
cannot be interfaced directly to an arithmetic coder;
arithmetic coders can typically only encode
(sequences of) discrete univariate choices.
%
One possible solution is to factorise~\eqref{eq:iid-ms}
into a product of univariate distributions as follows:
%
%
%
%
%
\def\mlt{<}%
\def\mle{\le}%
\def\bmlt{\text{\tiny$(\mlt)$}}%
\begin{equation}\label{eq:iid-ms-univar}
  \p{\setM \| N, \txtDD}
  =   \prod_{x \in \setX}^{\bmlt}
         \Binomial{ \mm{x} \| \MM\?-\sum_{y \mlt x} \mm{y},\ 
                              \frac{\DD{x}}{1\?-\sum_{y \mlt x} \DD{y}}
                  }
\end{equation}
where
$\Binomial{k \| K, \theta}$ $=$ ${K \choose k} \theta^{k} (1\?-\theta)^{K-k}$
is the binomial distribution,
$\mlt$ is a total ordering relation of the alphabet $\setX$,
and the product iterates over the elements of $\setX$ in the order
defined by $\mlt$.
A derivation of this factorisation can be found in~\thesisChapter{3}.

$\setM$ can then be compressed with an arithmetic coding scheme that
sequentially encodes
each of the components $\mm{x}$ using the
conditional distributions given in~\eqref{eq:iid-ms-univar}.
The length of the resulting compressed output sequence
%
%
is within 2 bits of $\setM$'s
information content, i.e.~the optimal length in bits:
\begin{eqnarray}\label{eq:ic-iid-multiset}
\log_2 \frac{1}{\p{\setM \| N, \txtDD}}
  &=& \log_2 \frac{1}{N!}
      + \sum_{\Mbot}^{\Mtop} \mk \log_2 \frac{1}{\DD{x}}
      + \sum_{\Mbot}^{\Mtop} \log_2 {\mk!}
\end{eqnarray}

\section{Permutations}\label{sec:perm}

A~\emph{permutation} is an arrangement of elements in a
given multiset.
For example, the multiset $\bset{\sym{X},\sym{Y},\sym{Z}}$
has 6~possible permutations:
\begin{displaymath}
\null\hfill
(\sym{X},\sym{Y},\sym{Z}) \qquad
(\sym{X},\sym{Z},\sym{Y}) \qquad
(\sym{Y},\sym{X},\sym{Z}) \qquad
(\sym{Y},\sym{Z},\sym{X}) \qquad
(\sym{Z},\sym{X},\sym{Y}) \qquad
(\sym{Z},\sym{Y},\sym{X}) \maybecomma
\end{displaymath}
\def\sA{\blacktriangledown}%
\def\sB{\blacktriangle}%
and the multiset $\bset{\sA,\sB,\sB}$ has three permutations:
$(\sA,\sB,\sB)$,
$(\sB,\sA,\sB)$ and
$(\sB,\sB,\sA)$.
Despite what the above notation might suggest,
a permutation contains no information
about symbols or their occurrence counts:
it merely specifies an arrangement of symbols
for a \emph{given} multiset.
A~sequence can always be separated into
a multiset and a permutation of that multiset.
%
%
%

\noindent
The number of possible permutations of a given multiset $\setM$ is exactly
\begin{equation}
  \frac{\MM!}{\prod_{\Mbot}^{\Mtop} \mk!}
  \maybecomma
\end{equation}
%
so a uniform distribution over permutations 
assigns to each permutation the probability:
\def\equniformperm{%
    \p{\vc{x}{1}{\MM} \| \setM}
  =   \frac{1}{\MM!}
    \mul \prod_{\Mbot}^{\Mtop} \mk!
}%
\begin{equation}\label{eq:uniform-perm}
    \equniformperm
    \maybedot
\end{equation}
%
%
%
%
This distribution~\eqref{eq:uniform-perm} can be factorised into
a product of univariate distributions: 
\begin{eqnarray}
\p{\vc{x}{1}{N} \| \setM}
  &=& \prod_{n=1}^{N} \p{x_n \| \setM,\ \vc{x}{1}{n-1}} \\
  &=& \prod_{n=1}^{N} \frac{\mm{x_n} - \fcnt{x_k=x_n}{k=1}{n-1}}{N - n + 1}
      \label{eq:perm-seq}
  \maybedot
\end{eqnarray}
%
One way to store such a permutation with an arithmetic coder
is to
encode each $x_n$ with the conditional distribution~\eqref{eq:perm-seq},
iterating over the indices $n$ of the permutation.
%
This procedure produces an output length that is within 2~bits of
the Shannon information content of the permutation:
%
\begin{eqnarray}\label{eq:ic-uniform-perm}
  \log_2 \frac{1}{ \p{\vc{x}{1}{\MM} \| \setM}}
  &=&
      \log_2 {\MM!}
    - \sum_{\Mbot}^{\Mtop} \log_2 \mk!
\end{eqnarray}

\section{Sequence = Multiset + Permutation}\label{sec:equality}

Any sequence $\vc{x}{1}{N}$ can be split into a multiset $\setM$
and a permutation of $\setM$.
%
For the case that each element $x_n$ is independent
and identically distributed according to some
known distribution~$\txtDD$,
we gave coding schemes for each of these three objects.
These coding schemes are optimal in the sense that they
compress each object to within 2 bits of the object's
Shannon information content.

The information content of the multiset was
given in~\eqref{eq:ic-iid-multiset},
and of the multiset's permutation in~\eqref{eq:ic-uniform-perm}.
Recalling that each symbol $x \in \setX$ occurs $\mm{x}$ times in
the sequence $\vc{x}{1}{N}$, and that $N = \MM$,
the sum of quantities \eqref{eq:ic-iid-multiset} and \eqref{eq:ic-uniform-perm} equals:
\begin{eqnarray}\label{eq:ic-ms-plus-perm}
       \sum_{\Mbot}^{\Mtop} \mm{x} \log_2 \frac{1}{\DD{x}}
  &=&  \sum_{n=1}^{N} \log_2 \frac{1}{\DD{x_n}}
  \maybedot
\end{eqnarray}
This quantity is the same as
expression~\eqref{eq:ic-iid-sequence},
the information content of the sequence.
%
%
%
%
%
%

\medskip

The above coding methods were derived for sequences and multisets
that made the oversimplifying
assumption that each element $x_n$ in the sequence was drawn
independently from a known distribution $\txtDD$.
In reality, distributions over elements are not typically
known in advance, and should be inferred from the data
(both by the compressor and by the decompressor).
Section~\ref{sec:learn} shows how sequences and multisets
can be optimally encoded while simultaneously learning
an unknown distribution. 

Before then, we shall briefly look at
truncated permutations,
and at two alternative models for multisets.
%
%

%

%

\section{Truncated permutations}\label{sec:truncperm}

The encoding for permutations introduced in section~\ref{sec:perm}
reflects a generative process where elements are drawn
without replacement from a given multiset $\setM$ until
there are no more elements.
In this section, we generalise this method to the case that
we stop after $K$ draws, resulting in a \emph{truncated permutation}
$\vc{x}{1}{K}$ where $K \le \MM$.

\def\Kpermutation{$K$\kern-.12em-per\-mu\-ta\-tion}%
The probability of drawing a given \Kpermutation\ from $\setM$ is:
\begin{equation}
  \p{\vc{x}{1}{K} \| K, \setM}
  = \frac{(\MM - \KK)!}{\MM!}
    \mul
    \prod_{x\in\setM} \frac{\mm{x}!}{(\mm{x}-\kk{x})!}
\end{equation}
where $\setK \subseteq \setM$ summarises the symbol
occurrence counts in the truncated permutation $\vc{x}{1}{K}$.
To encode a \Kpermutation\ of this kind, the encoding
procedure from section~\ref{sec:perm} can be used,
except for stopping after $K$ elements.
%
It might be helpful to note that this early stopping procedure
results in a distribution over truncated permutations that is non-uniform,
except when $K=\MM$ or when $\setM$ contains no duplicates.

\section{Combinations (or submultisets)}\label{sec:comb}

In section~\ref{sec:iid-ms} we considered multisets
whose elements were drawn from a distribution~$\txtDD$.
Rather than drawing elements from a known distribution~$\txtDD$,
consider drawing elements from a known multiset $\setM$
without replacement.
If we keep the order of the elements, we end up with
a permutation, as discussed in section~\ref{sec:perm}.
If the order is thrown away, the resulting structure is
called a \emph{combination}.
%
%
%

%
Suppose we draw $K$ elements from an existing multiset $\setM$,
where $K \le \MM$. The resulting combination $\setK$ is
a submultiset of $\setM$, and has probability:
\begin{equation}\label{eq:comb}
  \p{\setK \| \setM} 
  = {\MM \choose \KK}^{-1}
    \mul
    \prod_{x \in \setK} {\mm{x} \choose \kk{x}}
\end{equation}
This multivariate distribution can be broken down
into a product of univariate distributions, to
make it easier to use with an arithmetic coder.
For example, \eqref{eq:comb} can be factorised
recursively, using the identity:
\begin{equation}
\p{\setK \| \setM}
  \ =\
  \p{\kk{x} \| \setM, \KK}
  \mul
  \p{\setK_{\setminus \bset{x}} \|\big. \setM_{\setminus \bset{x}} }
\end{equation}
where $x$ is an arbitrary element that occurs in $\setK$,
and $\setK_{\setminus \bset{x}}$ denotes the multiset $\setK$ with all
occurrences of $x$ removed.
The distribution over $\kk{x}$ given $\setM$ and $\KK$ is:
\begin{equation}
  \p{\kk{x} \| \setM, \KK}
  \ =\
        {\MM \choose \KK}^{-1}
        {\MM\?-\mm{x} \choose \KK\?-\kk{x}}
        {\mm{x} \choose \kk{x}}
  \maybedot
\end{equation}

%
%
%

\section{Uniform multisets}\label{sec:uniform-ms}

The compression method for multisets described in section~\ref{sec:iid-ms}
assumes that the elements in the multiset are distributed
according to some distribution $\txtDD$.
\def\As{K}\def\as{k} 
\def\Cs{N}
\def\Csize{$N\kern-.1em$-size}%
%
%
This section shows how to compress
multisets that are chosen
uniformly at random from all possible multisets of a given size $\Cs$.
%
There are $C_{\As}(\Cs)$ ways of creating an \Csize\ multiset
from a set of $\As$ distinct elements, where:
\begin{equation}\label{eq:C_K}
C_{\As}(\Cs) = {\Cs+\As-1 \choose \As-1} = \frac{(\Cs+\As-1)!}{\Cs!\ (\As-1)!}
\end{equation}
A uniform distribution over multisets $\setM$ (given $\Cs$ and $\As$)
therefore assigns each multiset
probability mass $\p{\setM \| \Cs,\As} = 1 / {C_{\As}(\Cs)}$.
%
%

\section{Sequences and multisets from unknown distributions}\label{sec:learn}

Suppose the elements of a sequence $\vc{x}{1}{N}$ are independent
and identically distributed from some \emph{unknown}
distribution~$\txtDD$.
%
It is possible to compress sequences of this kind by taking
advantage of the knowledge that the sequence's elements
are identically distributed: even though the distribution
$\txtDD$ is not available to the encoder or the decoder,
each symbol in the sequence reveals information about
$\txtDD$, allowing both encoder and decoder to approximate it.

A simple and well-known technique is to let both encoder
and decoder maintain a histogram of symbol occurrence counts,
which is updated every time after a symbol in the sequence is
encoded or decoded.
Each symbol is arithmetically encoded using the best
approximation of $\txtDD$, given all the past symbol occurrences
and a prior distribution over symbols.
Here, this prior is implemented
by giving each symbol $x$ an initial pseudocount of~$\alpha_x \in \R^{+}$.
%
The $N$th symbol in the sequence can be compressed
with an arithmetic coder
using the following conditional distribution:
\def\hl#1{\colorbox{grey90}{$#1$}}%
\def\nohl#1{#1}%
\begin{eqnarray}
  \p{x_N\?=x \| \vc{x}{1}{N-1}, \v{\alpha}}
    &=& 
        \frac{\alpha_x + \hl{\fcnt{x_n=x}{n=1}{N-1}}} 
             {A + N - 1}
\end{eqnarray}
where $\alpha_x$ is the initial count for symbol $x$,
and the shaded box counts how often symbol $x$ occurred
in the preceding sequence
($\one{\,\mathrm Q\,}=1$ if
the expression $\mathrm Q$ is true, and $0$ otherwise).
$A$ is the sum of all $\alpha_x$.

The probability distribution over the entire sequence
induced by this adaptive scheme can be written as follows:
{\allowdisplaybreaks%
\begin{eqnarray}
  \p{\vc{x}{1}{N} \| N, \v{\alpha}}
     &=&  \prod_{n=1}^N \p{x_n \| \vc{x}{1}{n-1}, \v{\alpha} } \\
     &=&  \prod_{n=1}^N
           \frac{\alpha_{x_n} + \nohl{\fcnt{x_k=x_n}{k=1}{n-1}}}{A + n - 1}
             \label{eq:seq-learn-j1} \\
     &=&  \frac{\G{A}}{\G{A+N}}
          \prod_{x \in \setX}
          \frac{\G{\alpha_x + \mm{x}}}{\G{\alpha_x}}
             \label{eq:seq-learn-j2}
\end{eqnarray}%
}%
where $\setM$ is the symbol histogram of the entire sequence,
i.e.~$\mm{x} = \sum_{n=1}^N \one{x_n = x}$,
and $\G{\cdot}$ denotes the Gamma function.
%
%
%
%


\medskip

We can consider the equivalent scenario for a multiset
by throwing away the ordering produced by the above process.
As was shown in section~\ref{sec:perm},
a sequence with histogram $\setM$ has
exactly $\MM! \mul \left( \prod_{x\in\setX} \mm{x}! \right)^{-1}$
permutations.
Noting that $\MM = N$,
the distribution~\eqref{eq:seq-learn-j2} over sequences
can be converted to a distribution over multisets
by dividing out the permutation information:
%
%
\begin{equation}\label{eq:ms-learn}
    \p{\setM \| N, \v{\alpha} }
  = \frac{N!\ \G{A}}{\G{A+N}}
    \prod_{x \in \setX} \frac{\G{\mm{x}+\alpha_x}}{\mm{x}! \  \G{\alpha_x}}
\end{equation}
The result is a compound Dirichlet-multinomial distribution,
where $\v{\alpha}$ are the parameters of the Dirichlet prior.
The same result can be obtained by multiplying
distribution~\eqref{eq:iid-ms} with a Dirichlet prior,
and integrating out $\txtDD$.
%
%

To make this distribution suitable for compression
with an arithmetic coder,
it can be factorised into a product of conditional
Beta-binomial distributions, e.g.~as follows:
\begin{eqnarray}\label{eq:ms-learn-fact}
  \p{\setM \| N, \v{\alpha} }
  =
  \prod_{x \in \setX}^{\bmlt}
    \BetaBin{ \mm{x} \| 
                        N \?- \sum_{y<x} \mm{y},\ 
                        \alpha_x,\ 
                        A \?- \sum_{y \mlt x} \alpha_y
            }
\end{eqnarray}
where the Beta-binomial distribution is univariate, and defined as:
\begin{eqnarray}
\BetaBin{ k \| K, \alpha, \beta }
  &=& \int \Binomial{ k \| K, \theta }
      \mul \Beta{\theta \| \alpha, \beta} \  \d{\theta} \\
  &=& { K \choose k }
      \mul \frac{\G{\alpha\?+\beta}}{\G{\alpha} \  \G{\beta}}
      \mul \frac{\G{\alpha\?+k} \  \G{\beta\?+K\?-k}}{\G{\alpha\?+\beta\?+K}}
\end{eqnarray}
Using~\eqref{eq:ms-learn-fact} with an arithmetic coder,
one can near-optimally encode multisets whose elements
are distributed according to an unknown distribution.
It's somewhat surprising that it's possible to store
a collection of data points in an unordered way,
saving the bits that would otherwise be wasted
on the ordering. 
It's even more surprising that this is possible
while simultaneously learning the empirical distribution
from the data, and exploiting it to compress more
effectively.
%
%
Perhaps one could argue that this kind of compression method
is a step towards `universal' compression of unordered data.
\section{Discussion}\label{sec:discuss}

This paper presented several fundamental models
for non-sequential data,
and showed how they relate to some
known fundamental models for sequences.
The approach taken for finding these models
was to derive the probability distribution
over the combinatorial objects defined by
fairly basic assumptions.
Where necessary, multivariate distributions
were factorised into products of univariate
distributions, to make them usable with
existing arithmetic coding algorithms.

The models presented in this paper are not intended
to be used for compressing real-world data;
they were carefully chosen to be simplistic
and flexible, in the hope that they might serve as
fundamental components 
in more elaborate compression algorithms.

There are many other possibilities of defining
probability distributions over multisets,
and there are many other combinatorial objects
worthy of study that were not mentioned in this
paper.
Plenty of additional models and algorithms
related to compression of non-sequential data
can be found in~\citep{steinruecken2014b5}.

\ifx\isPreprint\undefined
\else
  \medskip

  \textsl{\textbf{Note.} This paper is a pre-print, to appear in the proceedings
  of the Data Compression Conference (DCC 2016).}

  \vspace*{-\medskipamount}%
\fi


%

\ifx\inDCC\undefined
\ifx\trimmedIEEE\undefined
  %
  %
\fi 
\fi 


%
%
%
\ifx\inArXiv\undefined
\fi
\ifx\inArXiv\undefined 
  \ifx\useIEEEbib\undefined
    \ifx\inDCC\undefined
      \bibliographystyle{mythesis-unsrt} 
    \else 
      \bibliographystyle{../style/IEEEtranN} 
      \section*{References}
    \fi
  \else
    \bibliographystyle{../style/IEEEtranN}  
  \fi
  \bibliography{cmpcmb}
\else 
  \bigskip
  \ifx\useIEEEbib\undefined
    \bibliographystyle{mythesis-unsrt}   
  \else
    \bibliographystyle{IEEEtranN}  
  \fi
  \bibliography{cmpcmb}
\fi
\end{document}